\documentclass[12pt]{article}
\usepackage{amssymb,amsmath,amsthm,graphicx,ulem}

\pdfoutput=1

\usepackage{graphicx,subfigure}
\usepackage{epsfig}
\usepackage{amsmath}
\usepackage{amsfonts}
\usepackage{amssymb}
\usepackage[usenames]{color}
\usepackage[letterpaper,left=2.cm,right=2.cm,top=2.5cm,bottom=2.5cm]{geometry}

\newcommand{\beq}{\begin{equation}}
\newcommand{\eeq}{\end{equation}}
\newcommand{\be}{\begin{equation}}
\newcommand{\ee}{\end{equation}}
\newcommand{\beqa}{\begin{eqnarray}}
\newcommand{\eeqa}{\end{eqnarray}}
\newcommand{\beqar}{\begin{eqnarray*}}
\newcommand{\eeqar}{\end{eqnarray*}}
\newcommand{\bea}{\begin{eqnarray}}
\newcommand{\eea}{\end{eqnarray}}

\numberwithin{equation}{section}

\newcommand{\nn}\nonumber
\newcommand{\eqn}[1]{(\ref{#1})}

\numberwithin{equation}{section}

\begin{document}

\allowdisplaybreaks

\normalem

\title{Optical Conductivity with Holographic Lattices}

\author{\\  Gary T. Horowitz${}^{\,a}$,
Jorge E. Santos${}^{\,a}$, David Tong${}^{\,b}$\\ 
\\ \\
  ${}^{\,a}$ Department of Physics, UCSB, Santa Barbara, CA 93106, USA \\ 
  ${}^{\,b}$ DAMTP,  University of Cambridge, Cambridge, CB3 0WA, UK \\
 \\
 \small{ gary@physics.ucsb.edu, jss55@physics.ucsb.edu, d.tong@damtp.cam.ac.uk}}
 
 \date{}

\maketitle

\begin{abstract}
\noindent  We add a gravitational background lattice to the simplest holographic model of matter at finite density and calculate the optical conductivity. With the lattice, the zero frequency delta function found in previous calculations (resulting from  translation invariance) is broadened and the DC conductivity is finite.  The optical conductivity exhibits a Drude peak with a cross-over to power-law behavior at higher frequencies. Surprisingly, these results bear a strong resemblance to the properties of some of the cuprates.
\end{abstract}

\newpage


\tableofcontents

\section{Introduction}

Over the past few years, it has been shown that various properties of condensed matter systems can be reproduced using general relativity \cite{Hartnoll:2009sz,Herzog:2009xv,McGreevy:2009xe,Sachdev:2010ch,sean2}. This surprising result is motivated by the remarkable gauge/gravity duality which relates a theory of gravity to a strongly coupled nongravitational theory. The duality is called ``holographic"  since the nongravitational theory lives in a lower dimensional space. 

The holographic approach has been applied to many interesting phenomena, including superconductivity and superfluidity, Fermi surfaces and non-Fermi liquids. However, with a few notable exceptions (see in particular \cite{Kachru:2009xf,Hellerman:2002qa}), much of this work has omitted one key ingredient of condensed matter systems: the lattice. 

The absence of an underlying lattice becomes particularly important in discussions of optical conductivity in systems with a finite density of charge carriers. The underlying translational invariance means that the charge carriers have nowhere to dissipate their momentum, resulting in a zero frequency delta function in the optical conductivity which obscures interesting questions such as the temperature dependence of the DC resistivity. Until now, the most common method to avoid this was to treat the charge carriers as probes, their momentum absorbed by the large number of neutral fields represented by the bulk geometry \cite{Karch:2007pd,Faulkner:2010da,Hartnoll:2009ns}.  

In Section 2 of this paper, we begin to remedy the situation by constructing a simple gravitational dual of a system with a lattice. We start with the minimal ingredients needed to describe a $2+1$ dimensional system at finite density: four dimensional gravity coupled to a Maxwell field. The equilibrium configuration at nonzero temperature is just a charged black hole. We then add the lattice by introducing a neutral scalar field with boundary conditions corresponding to a periodic source. This spatially varying field backreacts on the metric and gauge field, imprinting the lattice on the bulk geometry. We numerically solve the coupled Einstein-Maxwell-scalar field equations to find this gravitational crystal\footnote{To make the problem computationally manageable, we only add the lattice in one dimension and only compute conductivity in the direction of the lattice.}.  

By perturbing the lattice background, we can compute the optical and DC conductivities without the need to work in a probe approximation.  Normally, a perfect lattice with no impurities has infinite conductivity due to Bloch waves. However, our system naturally includes dissipation due to the black hole horizon and we see the expected broadening of the zero frequency delta function.

Our results are described in Section 3. They contain several surprises. At low frequency we find that both the real and imaginary parts of the optical conductivity are well described by the simple Drude form
\be
\sigma(\omega) = \frac{K\tau}{1-i\omega \tau}
\ee
where the constant $K$ and relaxation time $\tau$ are determined from the data.
However, at intermediate frequency the optical conductivity exhibits a cross-over to power-law behavior
\be
|\sigma(\omega)| =  \frac{B}{\omega^{2/3}} +C,
\ee
where $B$ and $C$ are constants. Rather strikingly, a power-law fall off with this exponent is  seen in the normal phase of some of the cuprates exhibiting high temperature superconductivity \cite{vanderMarel}. (The offset $C$ that we find is apparently not present in these materials.)  This result is robust against changes in the temperature,  lattice spacing, and ``strength" of our lattice. We do not have a deep understanding of why our numerical experiments on a  simple gravitational model reproduces results seen in real materials. 

At high frequency, the optical conductivity rises to a positive constant. This is different from the cuprates, but agrees with previous calculations without the lattice and is simply a property of conformal field theories in $2+1$ dimensions.

Finally, the presence of the lattice also renders the optical conductivity finite as the frequency $\omega\rightarrow 0$. This allows us to determine the temperature dependence of the DC resistivity, $\rho = 1/\sigma(0)$. We find that the DC resistivity is strongly dependent on the lattice spacing. Very similar behavior was seen recently by Hartnoll and Hofman who performed an analysis of an ionic lattice, created by a periodic variation in the chemical potential  \cite{Hartnoll:2012rj}. They observed a scaling behavior of the form $\rho \sim T^{2\Delta}$ where $\Delta$ is a complicated function of the lattice spacing. This scaling can be traced to the locally critical $AdS_2$ near horizon region of the Reissner-Nordstr\"om black hole, with the exponent $\Delta$ equal to the dimension of the operator dual to the charge density evaluated at the lattice wavenumber. Even though we introduce our lattice differently, we find that the DC resistivity precisely reproduces the scaling of \cite{Hartnoll:2012rj}  at low temperatures.

\section{\label{sec:model}A Holographic Lattice}
The minimal ingredients necessary to compute conductivity in a holographic framework are provided by Einstein-Maxwell theory in AdS${}_4$.  To this we add a neutral scalar field $\Phi$ which we will use to source the lattice. We work with the Lagrangian, 
\begin{equation}
S= \frac{1}{16 \pi G_N}\int d^4 x\,\sqrt{-g}\left[R+\frac{6}{L^2}-\frac{1}{2}F_{ab}F^{ab}-2\nabla_a \Phi \nabla^a \Phi-4 V(\Phi)\right],
\label{eq:action}
\end{equation}
where  $L$ is the AdS length scale and $F= \mathrm{d} A$. 
Our choice of potential corresponds to a massive scalar field with mass $m^2 = -2/L^2$,
\begin{equation}
V(\Phi) = -\frac{\Phi^2}{L^2}.
\end{equation}
The equations of motion derived from the action \eqn{eq:action} take the following form
\begin{subequations}
\begin{align}
&G_{ab}\equiv R_{ab}+\frac{3}{L^2} g_{ab}-2[\nabla_a \Phi \nabla_b \Phi-V(\Phi) g_{ab}]-\left(F_{ac}F_{b\phantom{c}}^{\phantom{b}c}-\frac{g_{ab}}{4}F_{cd}F^{cd}\right)=0,
\label{eq:einstein}
\\
&\nabla_a F^{a}_{\phantom{a}b}=0,
\label{eq:maxwell}
\\
&\Box \Phi-V'(\Phi)=0.
\label{eq:scalar}
\end{align}
\label{eqs:motion}
\end{subequations}
\indent Throughout the paper we shall only consider solutions that live in the Poincar\'e patch of AdS. We parametrize the holographic radial direction by the coordinate $z$ and impose boundary conditions which fix a conformal boundary metric at $z=0$ to be of the form,
\begin{equation}
\mathrm{d}s^2_\partial = -\mathrm{d}t^2+\mathrm{d}x^2+\mathrm{d}y^2.
\end{equation}

We will introduce the gravitational lattice background by providing a spatially inhomogeneous source for the neutral scalar field. Near the boundary, $\Phi$ takes the form
\be \Phi \rightarrow z\phi_1 + z^2\phi_2 + {\cal O}(z^3)\ee
According to the AdS/CFT correspondence, $\phi_1$ should be regarded as the source for the dimension two operator dual to $\phi$, say $\mathcal{O}_\phi$, while $\phi_2$ represents the expectation value  $\langle\mathcal{O}_\phi\rangle$. A general, inhomogeneous static solution is sourced by $\phi_1(x,y)$. However, here we will consider solutions that preserve translational invariance in the $y$ direction, with the lattice  varying only in the $x$ direction. We choose the source $\phi_1$ to be 
\begin{equation}
\phi_1(x) = A_0 \cos (k_0 x)\,.
\label{eq:source1}
\end{equation}
We will refer to $k_0$ as the lattice wavenumber and $A_0$ as its amplitude. $k_0$ is related to the lattice size $l$ in the usual way, $k_0=2\pi/l$. In the rest of this section,  we describe these gravitational lattices in more detail.

\subsection{\label{subsec:lattice}The Lattice}

Since our background is both static and translationally invariant in the $y$ direction, the solution is co-homogeneity two: it depends only on the coordinates $x$ and $z$. 
The most general static, electrically charged, black hole solution compatible with our symmetries can be written as
\begin{subequations}
\begin{equation}
\mathrm{d}s^2=\frac{L^2}{z^2}\left[-(1-z)P(z)Q_{tt}\mathrm{d}t^2+\frac{Q_{zz}\mathrm{d}z^2}{P(z)(1-z)}+Q_{xx} (\mathrm{d}x+z^2Q_{xz} \mathrm{d}z)^2+Q_{yy} \mathrm{d}y^2\right],
\label{eq:line}
\end{equation}
with
\begin{equation}
\Phi = z\,\phi(x,z)
\end{equation}
and
\begin{equation}
A = (1-z)\,\psi(x,z)\,\mathrm{d}t
\end{equation}
\label{eqs:ansatz}
\end{subequations}
where $Q_{ij}$, for $i\,j\in\{t,x,y,z\}$, $\phi$ and $\psi$ are arbitrary functions of $x$ and $z$, to be determined by solving Eqs.~(\ref{eqs:motion}). The line element shown above is invariant under reparametrizations of $x$ and $z$, \emph{i.e.} we have not yet fully specified our coordinate system. We shall address this issue later. The factor $(1-z)$ that appears both in the metric and gauge field ensures that Eqs.~(\ref{eqs:ansatz}) have a smooth non-extremal horizon located at $z=1$, provided that $Q_{tt}(x,1)=Q_{zz}(x,1)$ and $Q_{xx}$, $Q_{yy}$, $Q_{xz}$, $\psi$ and $\phi$ are smooth functions at $z=1$. Finally, the factor $P(z)$ is chosen to be
\begin{equation}
P(z) = 1 + z + z^2 -\frac{\mu_1^2 z^3}{2}
\end{equation}
and controls the black hole temperature
\begin{equation}
T= \frac{P(1)}{4\pi\,L}=\frac{6-\mu_1^2}{8\pi\,L}.\label{p1}
\end{equation}
Note that if $Q_{tt}=Q_{zz}=Q_{xx}=Q_{yy} = 1$, $Q_{xz}=\phi=0$, $\psi = \mu = \mu_1$, we recover the familiar planar Reissner-Nordstr\"om black hole.

Finding solutions to the system of equations (\ref{eqs:motion}), together with the ansatz (\ref{eqs:ansatz}), is not a well defined problem. The reason being that the Einstein equations do not have a definite character, in the PDE sense, if a gauge is not chosen. 
To overcome this difficulty we will use the DeTurck method, which was first outlined in \cite{Headrick:2009pv}. In this method we change the equations of motion by introducing suitable ``kinetic terms'' for $Q_{zz}$, $Q_{xx}$ and $Q_{xz}$, and ensure that any solution to this new set of equations is a solution to our initial problem, in a specific gauge.

\subsubsection{\label{subsubsec:turck}DeTurck method}
The DeTurck method is based on the so called Einstein-DeTurck equation, which can be obtained from Eq.~(\ref{eq:einstein}), by adding the following new term
\begin{equation}
G^{H}_{ab} \equiv G_{ab}-\nabla_{(a}\xi_{b)}=0,
\label{eq:einsteindeturck}
\end{equation}
where $\xi^a = g^{cd}[\Gamma^a_{cd}(g)-\bar{\Gamma}^a_{cd}(\bar{g})]$ and $\bar{\Gamma}(\bar{g})$ is the Levi-Civita connection associated with a reference metric $\bar{g}$. The reference metric is chosen to be such that it has the same asymptotics and horizon structure as $g$. For the case at hand, we choose $\bar{g}$ to be given by the line element (\ref{eq:line}) with $Q_{tt}=Q_{zz}=Q_{xx}=Q_{yy}=1$ and $Q_{xz}=0$. The DeTurck equation can be shown to be elliptic for a line element of the form (\ref{eq:line}) \cite{Headrick:2009pv}.

It is easy to show that any solution to $G_{ab}=0$ with $\xi=0$ is a solution to $G^{H}_{ab}=0$. However, the converse is not necessarily true. In certain circumstances one can show that solutions with $\xi\neq 0$, coined Ricci solitons, cannot exist \cite{Figueras:2011va}. Since the above equations are elliptic, they can be solved as a boundary value problem for well-posed boundary conditions and the solutions should be locally unique. This means that an Einstein solution cannot be arbitrarily close to a soliton solution and one should easily be able to distinguish the Einstein solutions of interest from solitons by monitoring $\xi$. In our numerical method we will solve Eq.~(\ref{eq:einsteindeturck}) directly, together with Eq.~(\ref{eq:maxwell}) and Eq.~(\ref{eq:scalar}), and a posteriori check that $\xi =0$, to machine precision. This will correspond to a solution of Eqs.~(\ref{eqs:motion}) in ``generalized'' harmonic coordinates defined by $\xi=0$.

In order to show that a solution can be found, we need to guarantee that suitable boundary conditions are given, and that these are compatible with $\xi=0$.

\subsubsection{\label{subsubsec:boundary}Boundary conditions}
We begin by discussing the boundary conditions close to the conformal boundary. For a solution of Eq.~(\ref{eq:einsteindeturck}) with $\xi=0$, one may compute the asymptotic behavior of the field content close to the conformal boundary
\begin{subequations}
\begin{align}
&Q_{tt}(x,z) = 1-\frac{z^2}{2}\phi_1(x)^2+\mathcal{O}(z^3)\,,
\\
&Q_{zz}(x,z) = 1+\frac{4z^2}{3}\phi_2(x)\phi_1(x)+\mathcal{O}(z^3)\,,
\\
&Q_{xz}(x,z)=\frac{z}{2}\phi_1(x)\phi'_1(x)+\mathcal{O}(z^2)\,,
\\
&Q_{xx}(x,z)=1-\frac{z^2}{2}\phi_1(x)^2+\mathcal{O}(z^3)\,,
\\
&Q_{yy}(x,z)=1-\frac{z^2}{2}\phi_1(x)^2+\mathcal{O}(z^3)\,,
\\
&\phi(x,z) = \phi_1(x)+z\phi_2(x)+\mathcal{O}(z^2)\,,
\\
&\psi(x,z) = \mu+[\mu-\tilde \rho(x)]z+\mathcal{O}(z^2),
\end{align}
\label{eqs:expansion}
\end{subequations}
where $\mu$ is a constant that denotes the chemical potential, $\tilde \rho(x)$ is the charge density and $\phi_i(x)$ are the two possible asymptotic decays for the neutral scalar field $\phi$. As explained in \eqn{eq:source1}, we introduce the lattice by choosing a source $\phi_1(x)$ with a nontrivial $x$ dependence of the form
\begin{equation}
\phi_1(x) = A_0 \cos (k_0 x)\,.
\label{eq:source}
\end{equation}
Notice that although the scalar field has a lattice wavenumber $k_0$, the stress-tensor is quadratic in $\Phi$, ensuring that the lattice will be imprinted on the metric and gauge field components $Q_{ij}$ and $\psi$ with  effective wavenumber of $2\, k_0$.

Using the expansion (\ref{eqs:expansion}) one can read off the boundary conditions at conformal infinity which, as expected, turn out to be of the Dirichlet type:
\begin{align}
&Q_{tt}(x,0)=Q_{zz}(x,0)=Q_{xx}(x,0)=Q_{yy}(x,0)=1,\nonumber
\\
&Q_{xz}(x,0)=0,\quad \phi(x,0)=\phi_1(x)\quad \psi(x,0) = \mu.
\end{align}
At the horizon ($z=1$), regularity demands the following expansion
\begin{subequations}
\begin{align}
&Q_{ij}(x,z)=Q_{ij}^{(0)}(x)+(1-z)Q_{ij}^{(1)}(x)+\mathcal{O}\left((1-z)^2\right)\,,
\\
&\phi(x,y)={\phi}^{(0)}(x)+(1-z){\phi}^{(1)}(x)+\mathcal{O}\left((1-z)^2\right)\,,
\\
&\psi(x,y)=\psi^{(0)}(x)+(1-z)\psi^{(1)}(x)+\mathcal{O}\left((1-z)^2\right)\,,
\end{align}
\end{subequations}
where the ${}^{(1)}$ terms can be determined in terms of the ${}^{(0)}$ terms and their tangential derivatives, by solving the equations of motion order by order in $(1-z)$. The boundary conditions are Dirichlet in $Q_{tt}$, because the above expansion demands $Q_{tt}(x,1)=Q_{zz}(x,1)$, and Robin boundary conditions for the remaining variables.

\subsubsection{\label{subsubsec:numerics}Numerical method}

We have used a standard pseudospectral collocation approximation in $z$, $x$ and solved the resulting non-linear algebraic equations using a  Newton-Raphson method. We represent the dependence in $z$ of all functions as a series in Chebyshev polynomials and the $x$-dependence as a Fourier series, so the $x$-direction is periodically identified. Our integration domain lives on a rectangular grid, $(x,z)\in (0,2\pi/k_0) \times (0,1)$.

\begin{figure}[t]
\centerline{
\includegraphics[width=0.5\textwidth]{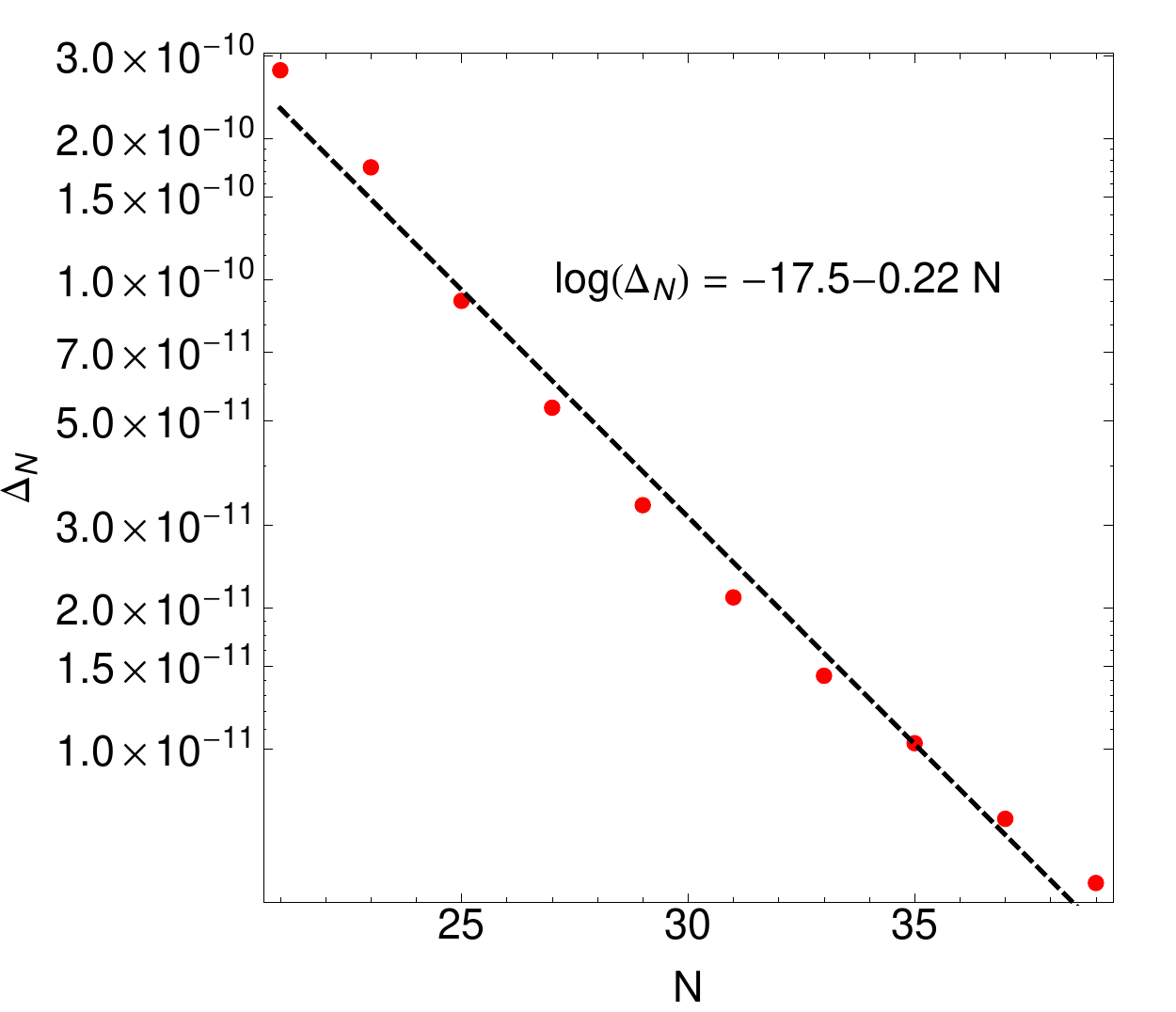}
}
\caption{$\Delta_N$ as a function of the number of grid points $N$. The vertical scale is logarithmic, and the data is well fit by an exponential decay: $\log(\Delta_N) = -17.5-0.22\,N$.}
\label{fig:convergence}
\end{figure}

In order to monitor the accuracy and convergence of our method, we have calculated the charge,  $\mathcal{Q}$, contained in our integration domain, for several resolutions. $\mathcal{Q}$ can be obtained by integrating $\tilde \rho(x)$ in the relevant range
\begin{equation}
\mathcal{Q}=\int_0^{2 \pi/k_0}\mathrm{d}x\,\tilde \rho(x).
\end{equation}
We denote the number of grid points in $x$ and $z$ by $N$ and compute $\Delta_N = |1-\mathcal{Q}_N/\mathcal{Q}_{N+1}|$ for several values of $N$. The results are plotted in Fig.~\ref{fig:convergence}. We find exponential convergence with increasing number of grid points, as expected for pseudospectral collocation methods.

Furthermore, in order to ensure that we are converging to an Einstein solution rather than a Ricci soliton we  monitor $\xi^a\xi_a$. All  plots shown in this manuscript have  $\xi^a\xi_a<10^{-10}$.

\subsubsection{Lattice background}

\begin{figure}[t]
\centerline{
\includegraphics[width=0.9\textwidth]{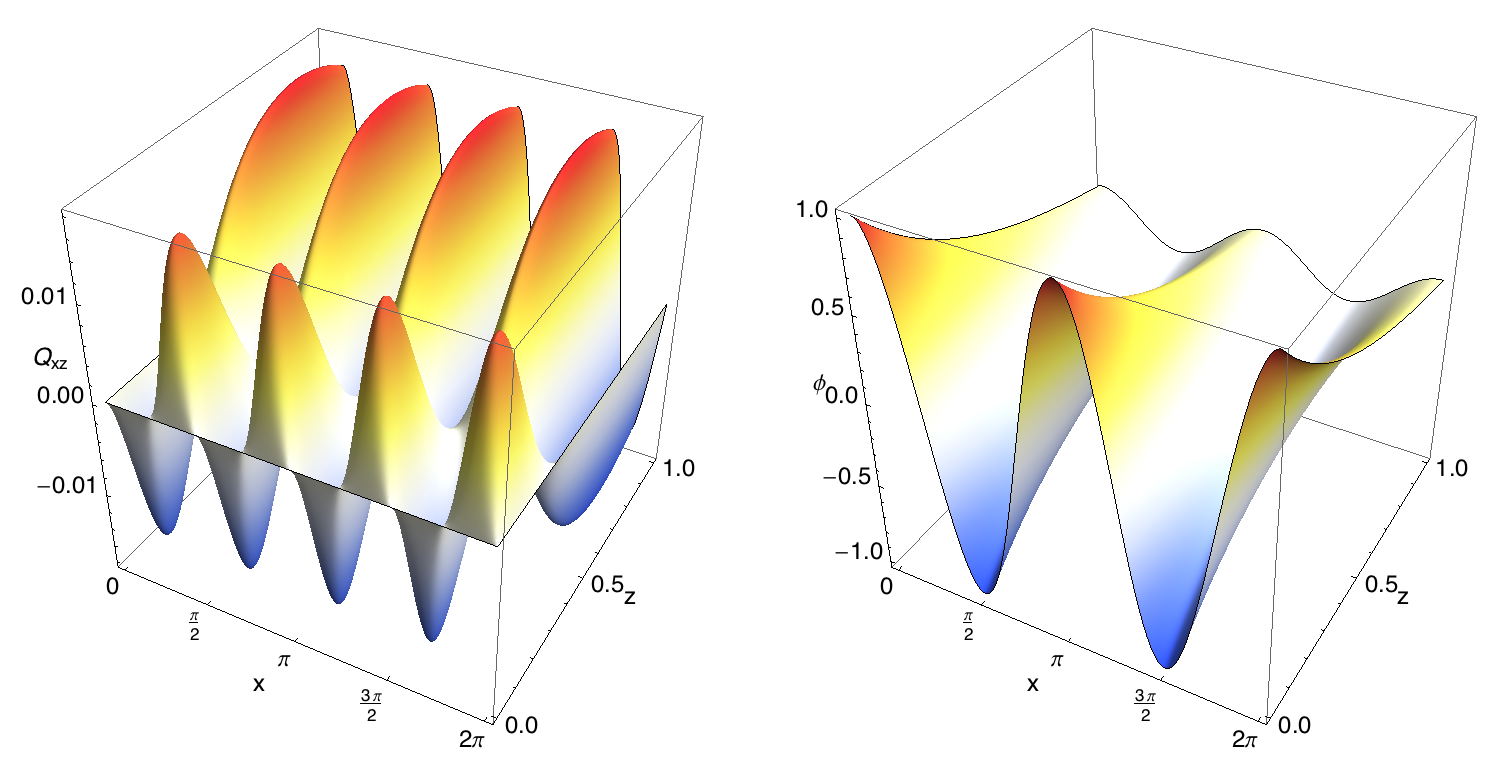}
}
\caption{On the left we show $Q_{xz}$ and on the right $\phi$, for $k_0=2$, $A_0 =1$, $\mu = 1.4$ and $T/\mu = 0.1$. Note that $Q_{xz}$ has an effective wavenumber of $2k_0$.}
\label{fig:typical}
\end{figure}

As an example, we show in Fig.~\ref{fig:typical} a typical result of our numerical code for $Q_{xz}$ and $\phi$. Note that the deviation in the metric due to the lattice, $Q_{xz}$,  is small even though the amplitude of the oscillating source generating the lattice is $A_0 = 1$.  It is important that this small metric correction is treated exactly and not just to first order,  since otherwise the perturbation that we will introduce to compute the conductivity would not feel the lattice. 

After solving the equations, we can study the dependence on temperature, lattice size, chemical potential and amplitude, by varying $\mu_1$, $k_0$, $\mu$ and $A_0$, respectively. Due to conformal invariance, we can work at fixed chemical potential and vary $T(\mu_1)/\mu$, $k_0/\mu$ and $A_0/\mu$, meaning that the physical moduli space is only three-dimensional.

Note that although the operator ${\cal O}_\phi$, dual to the scalar $\phi$, is relevant, the scalar perturbation decreases in the infrared. This is because the lattice source \eqn{eq:source1} is symmetric about $\phi=0$, while spatial gradients in the scalar field are  suppressed as we approach the horizon. Nonetheless, as is clear in Fig. \ref{fig:typical}, the scalar is not constant on the horizon.
In contrast, if we worked at $T=0$, the scalar field does indeed return to $\phi=0$ in the far infrared. In this sense, the lattice perturbation \eqn{eq:source1} is irrelevant.

\begin{figure}[t]
\centerline{
\includegraphics[width=0.9\textwidth]{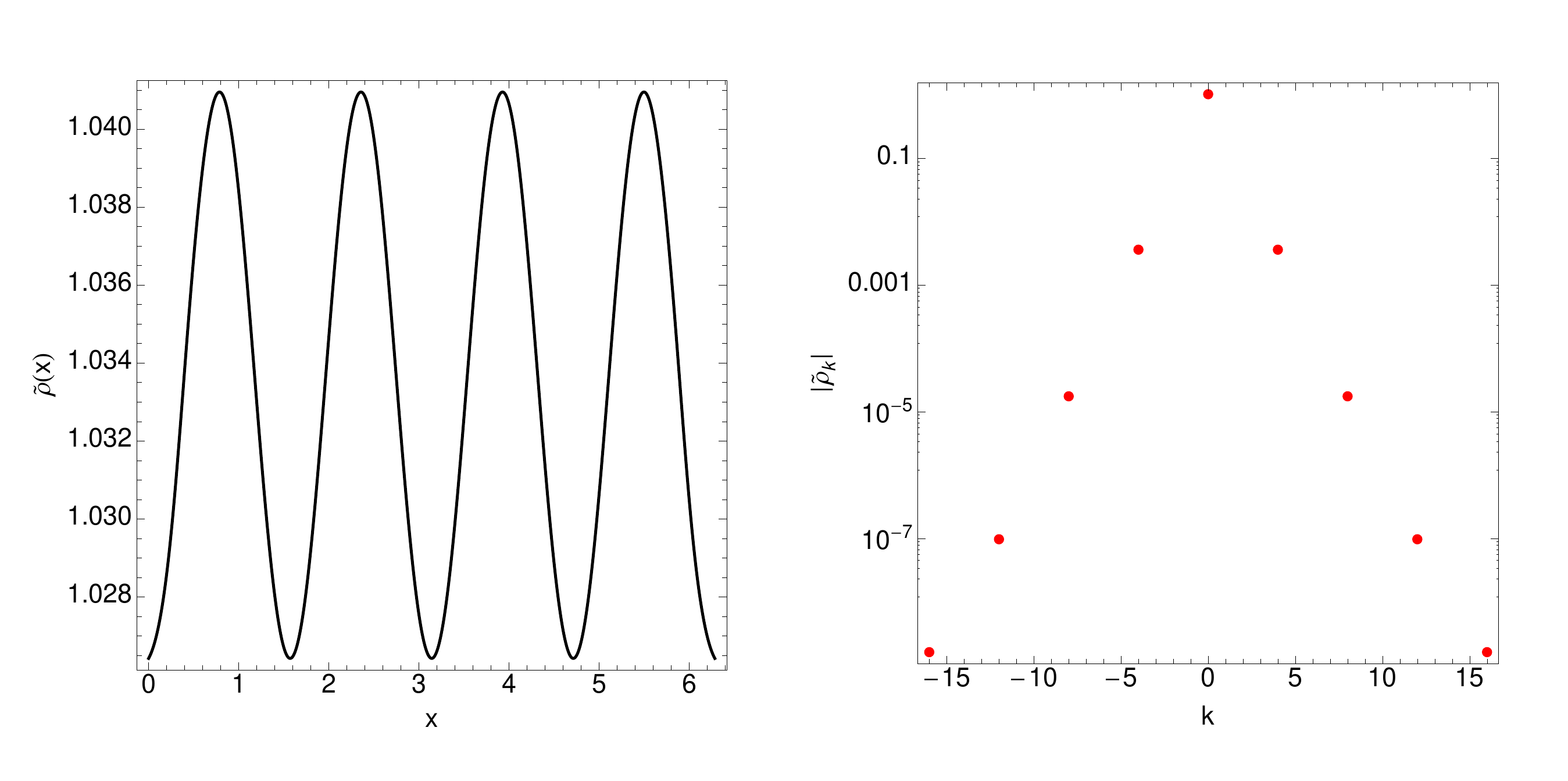}
}
\caption{On the left we show the charge density $\tilde \rho(x)$ and on the right the absolute value of the non-zero coefficients of its Fourier series.}
\label{fig:rho}
\end{figure}

Our lattices have $\tilde{\rho}/k_0^2\sim {\cal O}(1)$, which can be interpreted as order one charged degrees of freedom per lattice site\footnote{This is heuristic as we have only added the lattice in one direction.}. Although  
the lattice is sourced by a neutral scalar field, it nonetheless induces a mild spatial variation in the charge density 
 $\tilde \rho(x)$. On the left hand side of Fig.~\ref{fig:rho} we have plotted the charge density for a lattice with $A_0=1.5$, $k_0 = 2$, $\mu = 1.4$ and $\mu_1= 2.12$. This corresponds to a rather cold lattice with $T/\mu \simeq 0.043$. In order to better see the effects of the non-homogeneities caused by our choice of $\phi_1(x)$, we have decomposed $\tilde \rho(x)$ as a Fourier series, with coefficients $\tilde \rho_k$. On the right panel of Fig.~\ref{fig:rho} we show the resulting Fourier coefficients. Note that nonzero Fourier modes occur only for $k$ equal to multiples of $2k_0 = 4$. As mentioned earlier, this is because the stress tensor is quadratic in the scalar field, so the lattice seen by the metric and Maxwell field has twice the lattice wavenumber. For $N_x$ grid points along the $x$-direction, a Fourier grid can only hope to resolve up to $|k_{\max}|\leq N_x/2$. It is reassuring that for $k_0 = 2$, and with $N_x = 42$, the highest multipoles have a magnitude smaller than $10^{-10}$, meaning that our numerical code is capturing the relevant physics. Since higher wavenumbers come from higher powers of the scalar field in the nonlinear solution, the Fourier series exhibits an exponential decay with increasing wavenumber, which can be best seen in the logarithmic scale used on the right panel of Fig.~\ref{fig:rho}.

\subsection{\label{subsec:perturb}Perturbing the Lattice}

The main purpose of this paper is to explore transport properties in the presence of our lattice.  According to the AdS/CFT dictionary, on the gravity side, this is mapped into the study of perturbations about the lattice background.

Our first task is to write the equations governing generic perturbations of Eqs.~(\ref{eqs:motion}). We denote background fields with hats and expand all fields as
\begin{equation}
g_{ab}= \widehat{g}_{ab}+h_{ab},\quad A_a = \widehat{A}_a+b_a,\quad \Phi = \widehat{\Phi}+\eta\,,
\end{equation}
where $h_{ab}$, $b_a$ and $\eta$ should be regarded as small compared to $\widehat{g}_{ab}$, $\widehat{A}_{a}$ and $\widehat{\Phi}$, respectively. Expanding equations (\ref{eqs:motion}) to linear order determines how perturbations propagate in the background  $\{\widehat{g},\widehat{A},\widehat{\Phi}\}$. The resulting system of PDEs takes the following form
\begin{subequations}
\begin{multline}
\frac{1}{2}\left[-\widehat{\Box} h_{ab}-2 \widehat{R}_{acbd}h^{cd}+2 \widehat{R}_{(a\phantom{c}}^{\phantom{(a}c}h_{b)c}+2 \widehat{\nabla}_{(a}\widehat{\nabla}^c \bar{h}_{b)c}\right]= -\frac{3}{L^2}h_{ab}+4 \widehat{\nabla}_{(a}\eta \widehat{\nabla}_{b)}\widehat{\Phi}+2 V'(\widehat{\Phi})\eta\,\widehat{g}_{ab}+2 V(\widehat{\Phi})h_{ab}\\
+2 f_{(a\phantom{c}}^{\phantom{(a}c}\widehat{F}_{b)c}-\widehat{F}_{ac}\widehat{F}_{bd}h^{cd}-\frac{h_{ab}}{4}\widehat{F}^{cd}\widehat{F}_{cd}-\frac{\widehat{g}_{ab}}{2}f_{cd}\widehat{F}^{cd}+\frac{\widehat{g}_{ab}}{2}h^{cd}\widehat{F}_{cp}\widehat{F}_{d\phantom{p}}^{\phantom{d}p}
\label{eq:einsteinperturb}
\end{multline}
\begin{equation}
\widehat{\Box}b_a-\widehat{R}_{ac}b^c-\widehat{\nabla}_a(\widehat{\nabla}^c b_c)-\widehat{\nabla}_c\bar{h}^{cd}\widehat{F}_{da}-h^{cd}\widehat{\nabla}_c \widehat{F}_{da}-\widehat{F}^{cd}\widehat{\nabla}_{c}h_{da}=0
\label{eq:maxwellperturb}
\end{equation}
\begin{equation}
\widehat{\Box}\eta-\widehat{\nabla}_a\bar{h}^{ab}\widehat{\nabla}_{b}\widehat{\Phi}-h^{ab}\widehat{\nabla}_a\widehat{\nabla}_b \hat{\Phi}-V''(\widehat{\Phi})\eta = 0,
\label{eq:scalarperturb}
\end{equation}
\end{subequations}
where $f=\mathrm{d}b$, $\bar{h}_{ab} = h_{ab}-h\,\widehat{g}_{ab}/2$ is the so called trace-reversed metric perturbation, and curved brackets acting on indices indicate symmetrization.

The above system of equations is invariant under the following set of linear transformations
\begin{equation}
\begin{array}{l}
b_a \to b_a+\widehat{\nabla}_a \chi
\\
h_{ab}\to h_{ab}
\end{array}
\label{eq:gaugeele}
\end{equation}
and
\begin{equation}
\begin{array}{l}
b_a \to b_a +\beta^c\widehat{\nabla}_{c}\widehat{A}_a+\widehat{A}^d\widehat{\nabla}_d \beta_a
\\
h_{ab}\to h_{ab}+2\widehat{\nabla}_{(a}\beta_{b)},
\end{array}
\label{eq:gaugediffeo}
\end{equation}
where $\chi$ is an arbitrary scalar function and $\beta_a$ the components of an arbitrary four-dimensional vector.

The first set of transformations reflects the $U(1)$ gauge freedom associated with electromagnetism, whereas the second reflects diffeomorphism invariance. We shall fix these gauge freedoms by selecting the Lorentz and the de Donder gauge, respectively
\begin{equation}
\widehat{\nabla}^a \bar{h}_{ab}=0,\qquad \widehat{\nabla}^a b_a =0.
\label{eq:gaugefix}
\end{equation}
A few words about these gauge choices are in order. One can achieve any of these gauges by starting with an arbitrary metric and vector perturbations and then choosing $\chi$ and $\beta$ conveniently. This can always be done, because one can show that these two gauge conditions imply a wave-like equation for $\chi$ and $\beta$.

So far, our discussion about perturbation theory can be applied to any solution of \eqn{eqs:motion}: we  now specialize to (\ref{eqs:ansatz}). We want to study the changes in the transport properties induced by our lattice. Because our background is invariant under a time translation Killing field $\partial_t$, we can Fourier decompose our perturbations,
\begin{equation}
h_{ab}(t,x,y,z)=\tilde{h}_{ab}(x,y,z)e^{-i\omega t},\quad b_{a}(t,x,y,z)=\tilde{b}_{a}(x,y,z)e^{-i\omega t},\quad \eta(t,x,y,z) = \tilde{\eta}(x,y,z)\,e^{-i\omega t}.
\end{equation}
Given that the lattice extends along the $x$-direction, and leaves the $y$-direction unchanged, we will also assume translational symmetry along $y$. This in turn means that we will focus on perturbations with vanishing $\tilde{b}_y$, $\tilde{h}_{ty}$, $\tilde{h}_{zy}$ and $\tilde{h}_{xy}$, and that the remaining variables do not depend on $y$. We are left with $11$ unknown functions, $\{\tilde{h}_{tt},\tilde{h}_{tz},\tilde{h}_{tx},\tilde{h}_{zz},\tilde{h}_{zx},\tilde{h}_{xx},\tilde{h}_{yy},\tilde{b}_{t},\tilde{b}_{z},\tilde{b}_{x},\tilde{\eta}\}$ in two variables, $x$ and $z$.

It seems that the resulting system of equations governing perturbations about our lattice is overdetermined, because we have $15$ non-trivial differential equations to solve, namely the $\{tt,tz,tx,zz,zx,xx,yy\}$ components of Eq.~(\ref{eq:einsteinperturb}), the $\{t,z,x\}$ components of Eq.~(\ref{eq:maxwellperturb}), the scalar equation (\ref{eq:scalarperturb}), the $\{t,z,x\}$ components of the first equation in (\ref{eq:gaugefix}) and the last equation of (\ref{eq:gaugefix}). However, due to gauge invariance, the $\{tt,tz,yy\}$ components of Eq.~(\ref{eq:einsteinperturb}) and the $\{t\}$ component of Eq.~(\ref{eq:maxwellperturb}) are automatically satisfied if the remaining equations are satisfied. We have explicitly checked that this is indeed the case. We are left with a system of $11$ PDEs in eleven variables which can be solved using the numerical techniques of Sec.~\ref{subsubsec:numerics}. In the next section we shall present how to extract the relevant physics from the resulting numerical perturbations, and what boundary conditions do we choose in order to the study the lattice transport properties.

\subsubsection{\label{subset:bcsperturbations}Boundary conditions for the perturbations}

We wish to study the optical and DC conductivity of the dual CFT in the presence of the lattice. The DC conductivity can be extracted from the optical conductivity in the $\omega \to 0$ limit so, in this section, we focus on how to extract the optical conductivity only. This issue is naturally connected with what boundary conditions we impose for $\tilde{h}$, $\tilde{b}$ and $\tilde{\eta}$.

The optical conductivity of the boundary CFT is defined by the following
\begin{equation}
\tilde{\sigma}(\omega,x)\equiv\lim_{z\to 0}\frac{f_{zx}(x,z)}{f_{xt}(x,z)}.
\label{eq:opticalderiv}
\end{equation}
A couple of comments are in order about this expression. First, it is manifestly invariant under the local $U(1)$ of the electromagnetic field, because it is solely written in term of $f = db$ and not $b$ itself. Second, it is also gauge invariant under diffeomorphisms, because $f_{zx}$ is zero on the background solution (\ref{eqs:ansatz}). Note that $f_{zx}$ can be interpreted as a current, whereas $f_{xt}$ is an electric field. Third, the conductivity of the boundary field theory will in general be a function of $x$. Since we will impose a homogeneous boundary electric field, we are interested in the homogeneous part of the conductivity which we denote
as $\sigma(\omega)$. This is the quantity we study below. 

In order to proceed, we first need to generate a nonvanishing $f_{zx}$ and $f_{xt}$ on the boundary. This can be easily implemented by requiring $\tilde{b}_x$ to be a constant on the boundary, which without loss of generality we can set to $1$. This rescaling is always possible because we are studying a linear system of PDEs. A constant $\tilde{b}_x$ will generate a time dependent but homogeneous boundary electric field. We also do not want to change the boundary chemical potential, so we will set $\tilde{b}_t = 0$ at the boundary. Finally, we do not want to change the lattice, so we will demand $\tilde{\eta}$ to vanish as we approach $z=0$. For the remaining variables we demand normalizability at $z=0$. These boundary conditions induce the following asymptotic decays for $\tilde{b}$ and $\tilde{\eta}$:
\begin{equation}
\tilde{b}_t(x,z) = \mathcal{O}(z),\quad \tilde{b}_x(x,z) =1+j_x(x)\, z+\mathcal{O}(z^2), \quad \tilde{b}_z(x,z) =\mathcal{O}(z^2)\quad\text{and}\quad \tilde{\eta}(x,z)=\mathcal{O}(z^2).
\end{equation}
If we input these decays into (\ref{eq:opticalderiv}), we get the following simple expression for the optical conductivity in terms of boundary data only
\begin{equation}
\tilde{\sigma}(\omega,x) = \frac{j_x (x)}{i\,\omega}.
\end{equation}
For completeness, we also present the decays of the relevant metric perturbations as $z\to 0$
\begin{align}
& \tilde{h}_{tt}(x,z)=\mathcal{O}(z),\quad \tilde{h}_{tx}(x,z)=\mathcal{O}(z),\quad \tilde{h}_{tz}(x,z)=\mathcal{O}(z),\quad \tilde{h}_{xx}(x,z)=\mathcal{O}(z), \nonumber \\
& \tilde{h}_{xz}(x,z)=\mathcal{O}(z^4),\quad \tilde{h}_{zz}(x,z)=\mathcal{O}(z),\quad \text{and}\quad \tilde{h}_{yy}(x,z)=\mathcal{O}(z).
\end{align}

At the horizon, we impose ingoing boundary conditions, since this is required for causal propagation. The precise conditions on the perturbation are given in the   Appendix.

\begin{figure}
\centerline{
\includegraphics[width=0.9\textwidth]{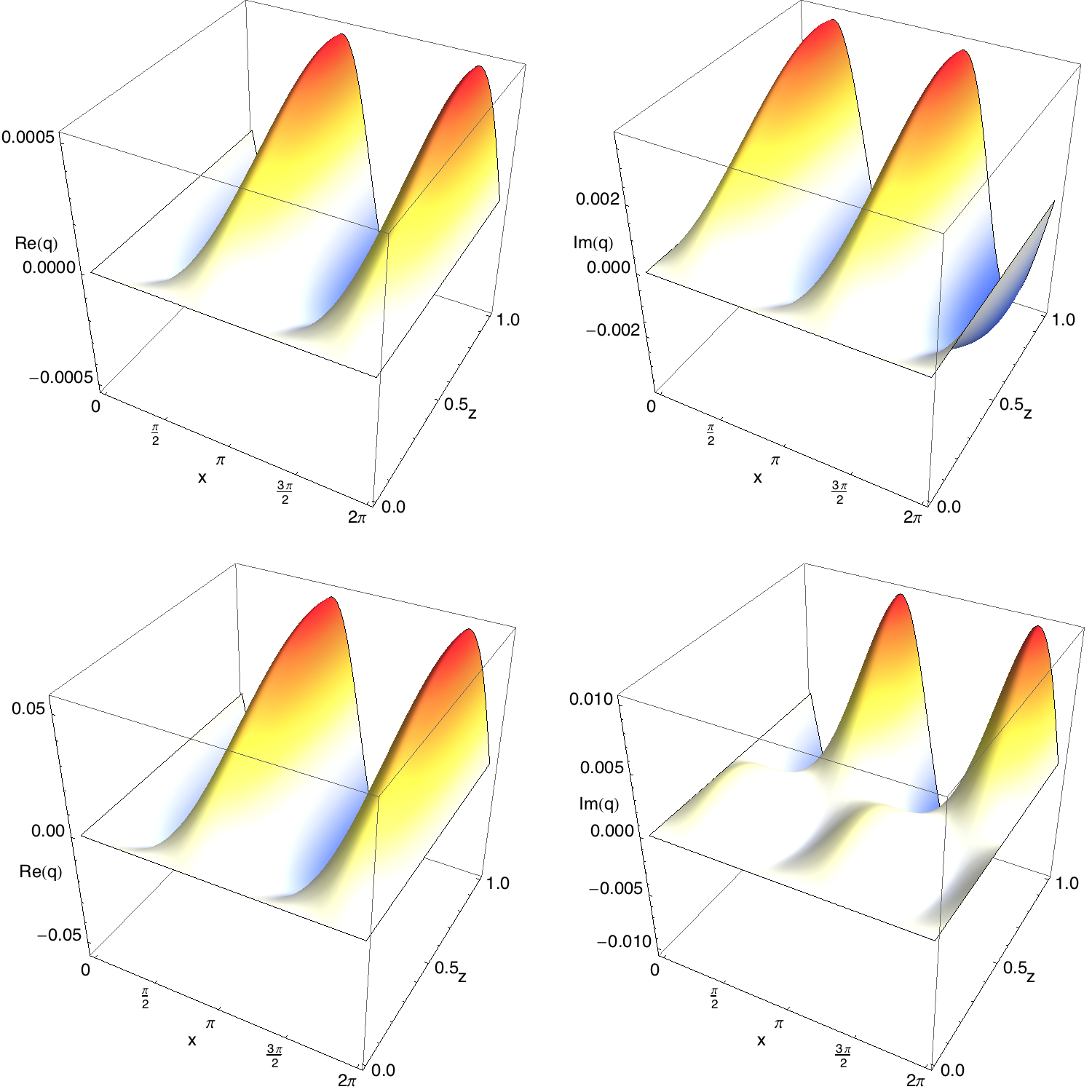}
}
\caption{The real and imaginary parts of the perturbation of the scalar field are shown for $\mu = 1.4$, $T/\mu = .115$, $k_0 = 2$, $A_0 = 1.5$. The two figures on the top have $\omega/T = 0.06$, whereas the two figures on the bottom have $\omega/T = 0.6$.}
\label{fig:typicalperturb}
\end{figure}

\subsubsection{\label{subset:numper}Numerical implementation of the perturbation theory}

Once we know the boundary conditions both at the horizon and asymptotic infinity, we just have to change to new variables which are adapted to the numerics. Because we are solving these equations using a pseudo-spectral collocation methods on a Chebyshev grid, we better guarantee that no non-analytic terms arise. In order to ensure that, we factor out the non-analytic behavior of the near horizon expansion (\ref{eqs:nearhexpansionper}) from the variables that we actually use in the numerics. For example, instead of using $\tilde{\eta}$ itself, we work with 
\begin{equation}
q(x,z) = z(1-z^3)^{\frac{i\,\omega}{P(1)}}\tilde{\eta}(x,z),
\label{eq:reno}
\end{equation}
with $P(1)$ given in \eqn{p1}. Note that at the conformal boundary $q(x,z)$ has a purely Dirichlet boundary condition $q(x,0)=0$ and at the horizon, because the field equations fix $\tilde{\eta}^{(1)}(z)$ in terms of the $\,^{(0)}$ coefficients, $q(x,z)$ has a Robin-type boundary condition. It turns out that we can always recast the boundary conditions at the conformal boundary and horizon as homogenous Dirichlet or homogeneous Robin boundary condition, by multiplying suitable powers of $z$ and $(1-z^3)^{\frac{i\,\omega}{3\,P(1)}}$ to our original variables. The only exception to this procedure is $\tilde{b}_x$, which has an inhomogenous Dirichlet boundary condition at $z=0$. Recall that we need $\tilde{b}_x(x,z)=1$ in order to generate a boundary electric field.

After discretization, our system of PDEs can be written as a linear map of the form
\begin{equation}
\bf M \cdot \bf x = \bf x^b,
\end{equation}
where $\bf x^b$ includes the nonhomogeneous boundary condition for $\tilde{b}_x$. This equation can then be solved by the \verb LinearSolve  in-built function in \emph{Mathematica}. In Fig.~\ref{fig:typicalperturb} we show the typical output of our code for $q(x,z)$ defined in Eq.~(\ref{eq:reno}).

\section{Conductivity}

Before describing our new results, we first review the  conductivity in a translationally invariant holographic background. For boundary theories with two spatial dimensions, the conductivity is dimensionless and at the conformal point, with $\mu=0$, is known to be a independent of $\omega$, reflecting an underlying electron-vortex duality \cite{hkss}.

\begin{figure}[t]
\centerline{
\includegraphics[width=0.9\textwidth]{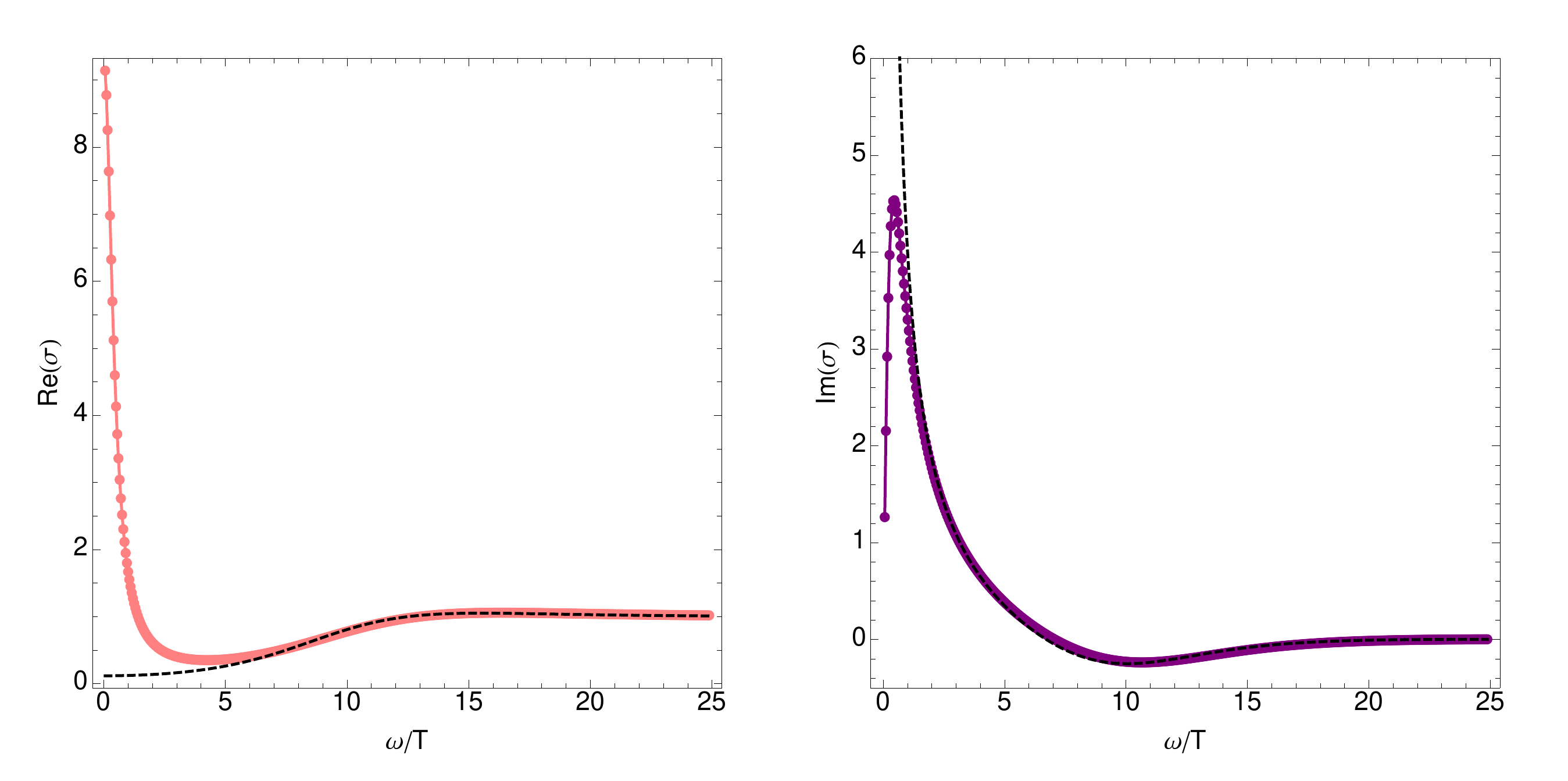}
}
\caption{The optical conductivity, both without the lattice (dashed line) and with the lattice (solid line and data points) for $\mu =1.4$ and temperature $T/\mu = 0.115$.  Note that the lattice (which has wavenumber $k_0=2$ and amplitude $A_0=1.5$) only changes the low frequency behavior.  The pole in ${\rm Im}\,\sigma$ without the lattice reflects the existence of a $\omega=0$ delta-function in ${\rm Re}\,\sigma$.}
\label{config}
\end{figure}

In the presence of a chemical potential, $\mu$, the optical conductivity shows more structure  \cite{Hartnoll:2009sz}. Both real and imaginary parts are shown by the dashed, black curves  in Fig.~\ref{config} for a temperature $T/\mu = 0.115$. 
At large frequency, $\omega \gg \mu$, the conductivity  tends towards the constant, real value observed at the conformal point. Indeed, this is the expected behaviour  for any scale invariant theory in two spatial dimensions. At  lower frequencies, $\omega < \mu$, a drop in ${\rm Re}\,\sigma$ reveals a depletion in the density of charged states.   However, for the purposes of our present discussion, the most important feature of the conductivity does not show up in numerical plots of ${\rm Re}\,\sigma$: it is a delta-function spike at $\omega =0$. 
The presence of this delta-function can be seen in the plot of ${\rm Im}\,\sigma(\omega)$ where, via the Kramers-Kronig relation, it reveals itself as a pole,
\be {\rm Im}\,\sigma(\omega)\rightarrow \frac{K}{\omega}\ \ \ \ \ {\rm as}\ \omega\rightarrow 0
\label{Kpole}\ee
for some constant $K$.

There is nothing mysterious about the presence of this delta-function. It follows solely on the grounds of momentum conservation in the boundary theory. If we have a translationally invariant state with nonzero charge density, then one can always boost it to obtain a nonzero current with zero applied electric field.  This results in the infinite DC conductivity.

The introduction of a background spatial lattice, as described in the previous section, resolves this issue.  With no translational invariance, there is no momentum conservation and the $\omega=0$ delta-function spreads out, revealing its secrets.    In this section we describe what  was hiding in that delta-function.

The  optical conductivity,  $\sigma(\omega)$, in the presence of the lattice is  shown by the solid line in Fig.~\ref{config}.  At high frequencies, $\omega \gg \mu$, the optical conductivity in the lattice background remains unchanged from the translationally invariant black hole. The interesting physics lies at lower frequencies. The dissipative part of the conductivity, ${\rm Re}\,\sigma$, now rises at low $\omega$. This is the redistribution of  the spectral weight of the delta-function. Moreover, the pole in the responsive part of the conductivity, ${\rm Im}\,\sigma$, has now disappeared, with   ${\rm Im}\,\sigma(\omega)\rightarrow 0$, as $\omega\rightarrow 0$, confirming that  there is no longer a delta-function at zero frequency\footnote{The resolution of a delta-function into a Drude-like peak has been seen in a somewhat different context in conformal fixed points with vanishing charge density \cite{damle}. Here the delta-function is resolved by interactions rather than breaking of translational symmetry, either in an $\epsilon$ expansion \cite{damle} or a $1/N$ expansion \cite{sachdev}.}.  We now describe the characteristics of the conductivity in more detail.

\subsection{The Drude Peak}

\begin{figure}[t]
\centerline{
\includegraphics[width=0.9\textwidth]{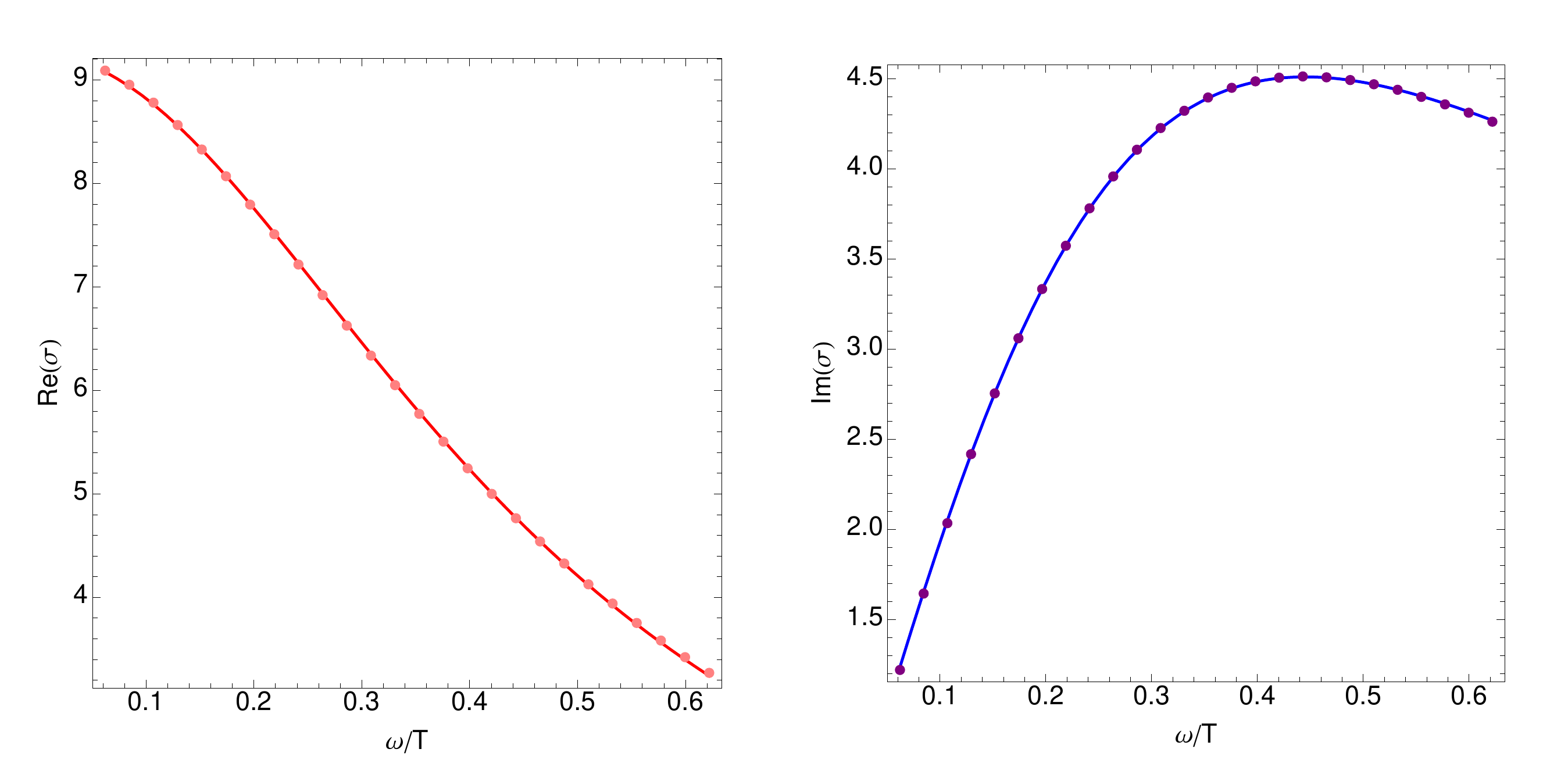}
}
\caption{A blow up of the low frequency optical conductivity with lattice shown in Fig. \ref{config}. The data points in both curves are fit by the simple two-parameter Drude form \eqn{drude}. }
\label{drudefig}
\end{figure}

At low frequency, both the real and imaginary parts of the conductivity can be fit by the two-parameter Drude form
\be \sigma(\omega) =\frac{K \tau}{1-i\omega\tau}\label{drude}\ee
with both the scattering time $\tau$ and the  overall amplitude $K$ constants, independent of $\omega$. This is shown in Fig. 
 \ref{drudefig}.  It can be checked numerically that the overall amplitude $K$ agrees (to about the $1\%$ level) with the coefficient of the pole  \eqn{Kpole} in the translationally invariant case. All interesting physics in this regime is therefore captured by the single parameter, $\tau$.  We have varied the temperature and lattice spacing and found that this Drude form holds in all cases. Given the lack of well-defined quasi-particles in our holographic system, it seems surprising that the low-frequency behaviour of our system is governed so well by the exact Drude form.

\begin{figure}[t]
\centerline{
\includegraphics[width=1.0\textwidth]{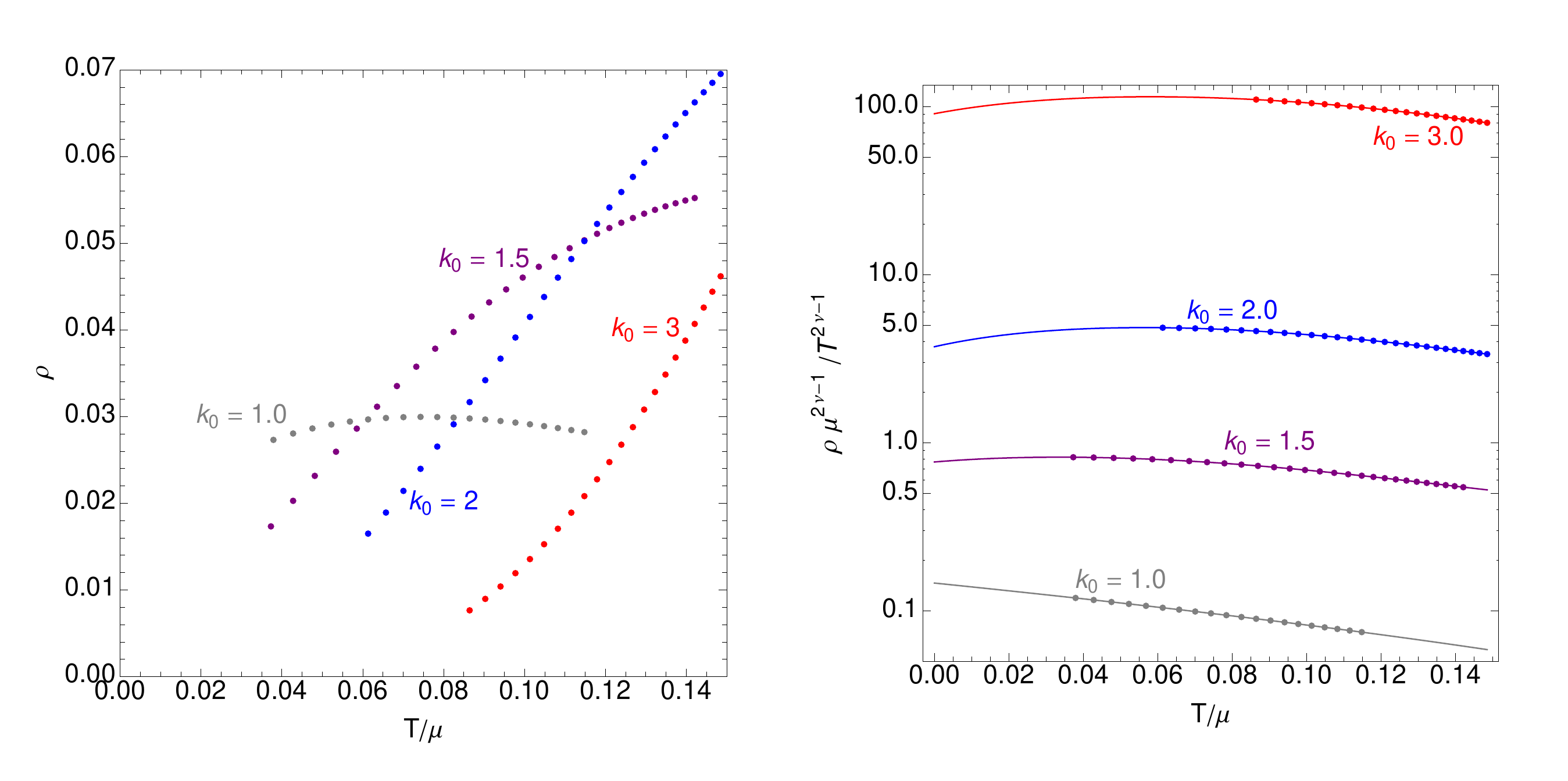}
}
\caption{The left panel shows the  DC resistivity plotted as a function of temperature  for various lattice spacings. On the right hand side we factor out the scaling (\ref{Tscaling}, \ref{nu}) and re-plot the same data on a log scale. The lines denote a fit to the data including polynomial corrections to the leading low temperature behavior.
Both plots arise from a background with $\mu =1.4$ and the lattice amplitude $A_0 = k_0/2$. The plots remain essentially unchanged for lattices of different amplitudes.}
\label{dcfig}
\end{figure}

\subsection{DC Resistivity}

The resolution of the $\omega=0$ delta-function leaves behind a well-defined DC resistivity, $\rho = (K\tau)^{-1}$.  
The Drude amplitude $K$  is essentially independent of temperature $T$ and all temperature dependence in the resistivity  $\rho(T)$ is inherited from $\tau$. The results depend strongly on the lattice wavenumber $k_0$ and are shown on the left hand side of Fig. \ref{dcfig}. 

To make sense of this complicated plot, we review some recent work in the literature.
Since the near horizon geometry of an extremal Reissner-Nordstr\"om AdS black hole is $AdS_2\times R^2$, the dual theory is said to be ``locally critical" in the sense that it is invariant under rescalings of time, with no rescaling of space. Hartnoll and Hofman \cite{Hartnoll:2012rj} have recently studied the DC conductivity in a locally critical theory. They showed that the DC conductivity can be extracted from the two point function of the charge density, evaluated at the lattice wavenumber. 
They then calculated this two point function by perturbing the Reissner-Nordstr\"om
AdS black hole and found
%
\be\label{Tscaling}
\rho \propto T^{2\nu -1}
\ee
where\footnote{This is a manifestly scale invariant form of the exponent that was found in \cite{Hartnoll:2012rj} and was first derived in  a different context in \cite{Edalati:2010pn}.}
\be\label{nu}
\nu =  \frac{1}{2}\sqrt{5+2(k/\mu)^2 - 4\sqrt{1 + (k/\mu)^2}}
\ee
The exponent  can be viewed as arising from the dimension $\Delta = \nu -\frac{1}{2}$ of the operator dual to the charge density in the near horizon $AdS_2$ region, evaluated at the lattice wavenumber $k$.

On the right hand side of  Fig. \ref{dcfig} we plot $\rho/T^{2\nu-1}$ for several values of the lattice wavenumber. As discussed earlier, if our scalar field has lattice wavenumber $k_0$, the charge density has lattice wavenumber $2k_0$, so we have set $k = 2k_0$ in (\ref{nu}). We have fit the data to $\rho_0 = T^{2\nu -1}(a_0 + a_1 T + a_2 T^2 + a_3 T^3)$ and drawn the curves on the right hand side of  Fig. \ref{dcfig}. The fact that the curves all  approach nonzero, but finite,   constants at low temperature shows that our  data confirms the low temperature scaling (\ref{Tscaling}) with exponent (\ref{nu}) predicted in \cite{Hartnoll:2012rj}. 

Note that as the temperature goes to zero, the dissipation goes to zero and the DC resistivity vanishes. Thus the DC conductivity becomes infinite, as expected for a perfect lattice with no dissipation.

\begin{figure}[t]
\centerline{
\includegraphics[width=0.9\textwidth]{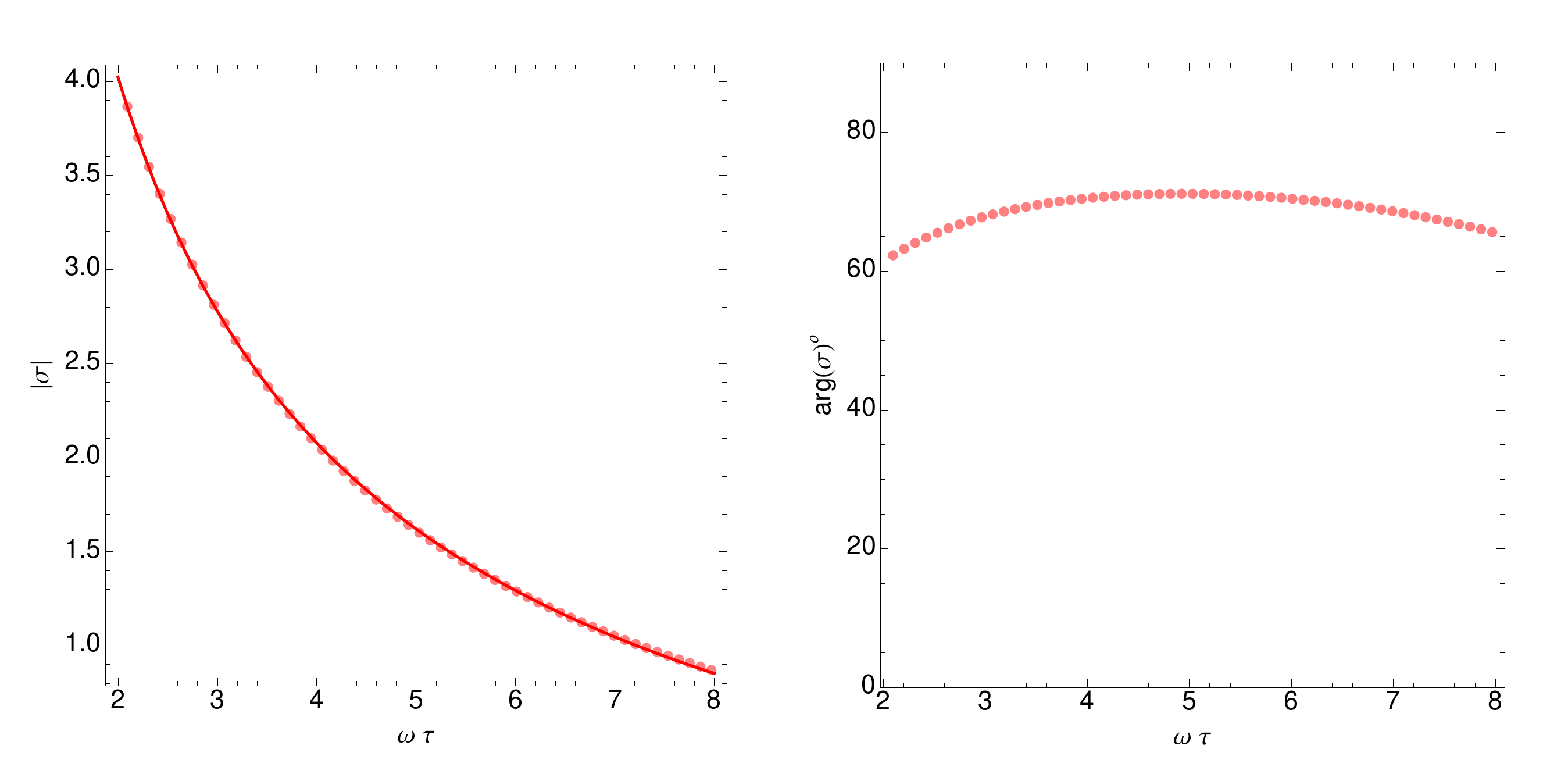}
}
\caption{The modulus $|\sigma|$ and argument ${\rm arg}\,\sigma$ of the conductivity. The background for both plots has wavenumber $k_0=2$, amplitude $A_0=1.5$, chemical potential $\mu =1.4$ and temperature $T/\mu = 0.115$. The line on the right is a fit to the power law \eqn{power}.}
\label{powerfig}
\end{figure}

\subsection{Power-law Optical Conductivity}

The fit to the Drude peak works well for $\omega /T\lesssim 1$. However, for $ \omega /T \gtrsim 1$,  the optical conductivity exhibits a power-law fall-off in a ``mid-infrared" regime, before reverting to the continuum result. It is convenient to use $\omega \tau$ as a dimensionless measure of the frequency in this region. This is because the power law behavior falls roughly in the range $2 < \omega \tau < 8 $ for all the lattices we have examined, even those with different temperature, lattice spacing and amplitude. (For the temperature we  use in Fig. 5, $\tau T = 2.22$, so $\omega \tau = 2$ corresponds to $\omega/T =  0.9$.) In Fig.~\ref{powerfig}  we have plotted $|\sigma|$ and the phase angle over this range of frequencies. The data is very well fit by
\be |\sigma(\omega)|  = \frac{B}{\omega^{2/3}} + C\label{power}\ee
Meanwhile the phase angle has only small dependence on $\omega$, but varies between $65^\circ$ and $80^\circ$, as $k_0$ varies from $1$ to $3$.
The slight variation in the phase angle is enough so that the real and imaginary parts of the conductivity do not individually follow simple power laws over the range indicated in Fig. \ref{powerfig}.   

\begin{figure}[t]
\centerline{
\includegraphics[width=1.0\textwidth]{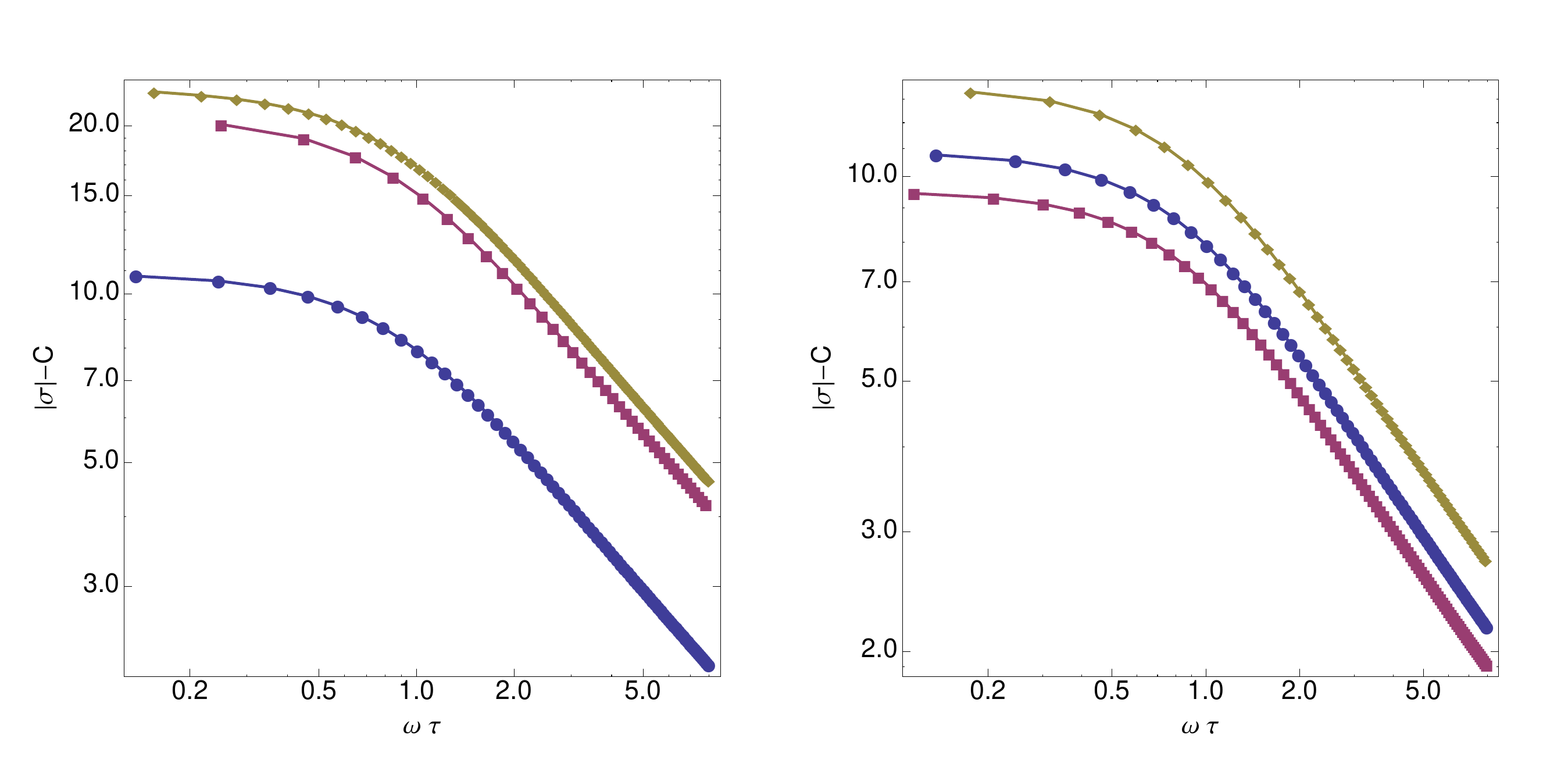}
}
\caption{The magnitude of the optical conductivity with the offset removed, on a log-log plot. On the left, the plot has $T/\mu = .115$,  and shows three different wavenumbers: diamonds denote $k_0 = 3$, the squares denote $k_0 = 1$, and the circles denote $k_0 = 2$. On the right, the plot has $k_0 = 2$ and shows three different temperatures: the diamonds have $T/\mu = .098$, the circles have $T/\mu = .115$ and the squares have $T/\mu = .13$.  In both plots, $A_0/k_0 = 3/4$. The fact that the lines are parallel for $\omega \tau >2$ shows that the fit to the power law \eqn{power}  is robust.}
\label{loggy}
\end{figure}

The exponent of the power law does not depend on the particular choices we have made for the parameters in our model. To illustrate this, and to make the power law more manifest, in Fig. \ref{loggy} we  show $(|\sigma|-C)$ vs $\omega\tau$ on a log-log plot. On the left we show three different choices for the lattice wavenumber $k_0$. On the right, we show three different temperatures. The fact that the curves  all form parallel straight lines  for $\omega \tau > 2$ shows the power law fall-off with exponent $-2/3$ is robust. Since the offset $C$ depends on $k_0$ and $T$, in Fig. \ref{loggy} we have subtracted a different constant for each curve.

\begin{figure}[t]
\centerline{
\includegraphics[width=0.9\textwidth]{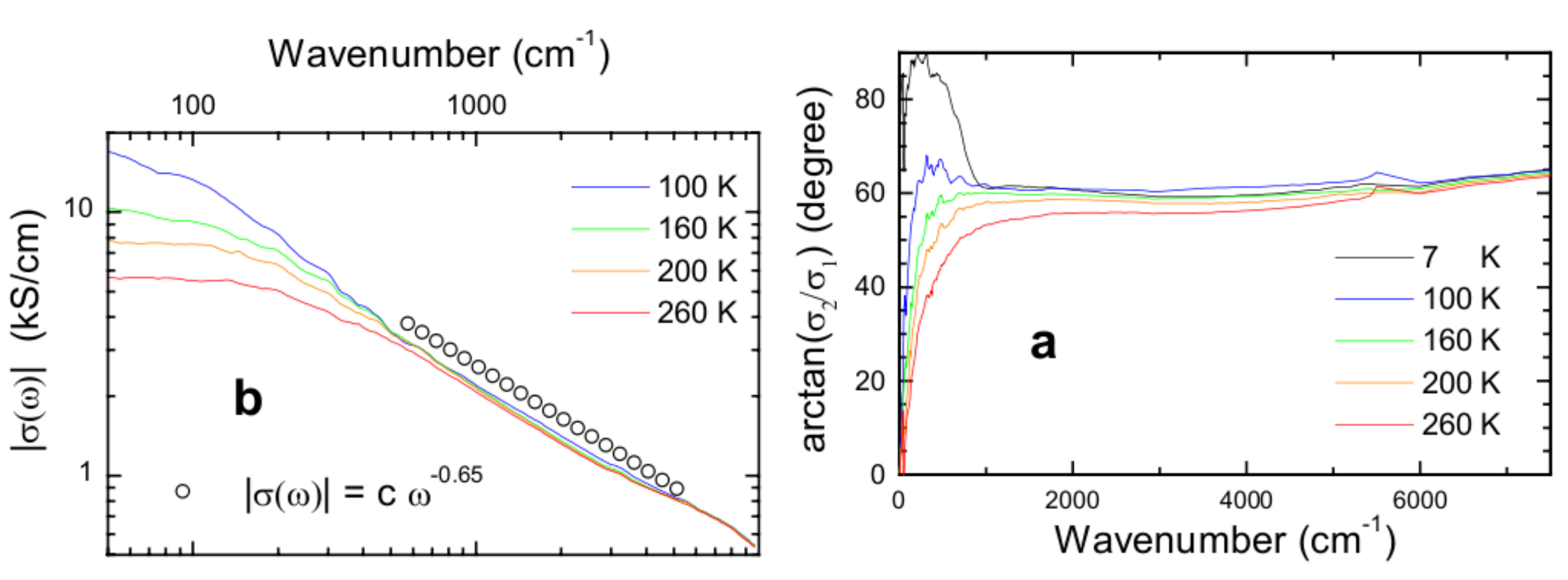}
}over
\caption{The optical conductivity of optimally doped $Bi_2Sr_2Ca_{0.92}Y_{0.08}Cu_2O_{8+\delta}$. This plot is taken from \cite{vanderMarel}.}
\label{nature}
\end{figure}

\subsection{Comparison to the Cuprates}

Our results for the optical conductivity bear a striking resemblance to the behaviour seen in the cuprates where a mid-infrared power-law contribution  has  long been observed \cite{azrak,vanderMarel,marel2}. At low frequencies, $\omega/T \lesssim 1.5$, the optical conductivity takes the Drude form \eqn{drude} if one also allows for $\tau$ to scale linearly with both $T$ and $\omega$ \cite{marel2}.  But for $\omega/T\gtrsim 1.5$, a cross-over to power-law behaviour takes place, in which $|\sigma|\sim\omega^{-\gamma}$  with $\gamma\approx 0.65$ and the phase is roughly constant around $60^\circ$ \cite{vanderMarel}. This is shown in Fig.~\ref{nature}.

There are differences between our behaviour and that of the cuprates. Most notably, the DC resistivity that we obtain is nothing like the robust linear behavior characteristic of the strange metal regime; instead we find power-law behaviour with an exponent that depends on the lattice wavenumber. Moreover, in the mid-infra red regime our optical conductivity requires a constant off-set $C$ in \eqn{power}. No such off-set  is seen in \cite{vanderMarel}. 

In some sense, this discrepancy makes the agreement of the phase angle somewhat more surprising. For strictly power-law conductivity (i.e. with $C=0$ as seen in the cuprates), causality and time-reversal invariance in the form $\sigma(\omega)=\sigma^\star(-\omega)$ relate the exponent of the power law to the phase \cite{vanderMarel}. However, with $C\neq 0$, there is no reason for the phase and exponent to be related. 

There have been previous theoretical observations of power-law fall-off in the optical conductivity. It is seen in  models of charges moving in a periodic potential subject to dissipation \cite{Kato}, but the exponent typically depends on the details of the model of dissipation. In the context of the cuprates, Anderson suggested a power-law fall-off $\sigma(\omega)\sim \omega^{-\gamma}$ on the basis of a Luttinger liquid model, with $\gamma=2/3$ arising from a coupling to a gauge field \cite{anderson}.  

Perhaps more pertinent for the present discussion, the universal power law observed in the optical conductivity, together with a $\omega/T$ scaling, was associated to an underlying quantum critical point in \cite{vanderMarel}. Of course the starting point of our holographic model is a strongly interacting critical point, albeit with a scale introduced by the finite density. Moreover, for small temperatures, $T\ll \mu$, our model exhibits an emergent  locally critical point, reflected by the near horizon  $AdS_2\times {\bf R}^2$ regime. 

However, a second explanation was put forward in \cite{norman} where it was argued that the $\sigma \sim \omega^{-\gamma}$ behavior with $\gamma\approx 0.65$ was a generic prediction of electrons interacting with a broad spectrum of bosons. Our holographic model is certainly not short of such bosonic modes and it is possible that these are responsible for our observed behavior. 

Finally, within the holographic framework, a $\sigma(\omega)\sim\omega^{-2/3}$ power-law was shown to arise from probe charged matter interacting with a strongly coupled soup with dynamical exponent $z=3$ \cite{Hartnoll:2009ns}.

\section{Future Directions}

The introduction of a gravitational lattice in the simplest holographic model of a conductor has allowed us to explore the low-frequency optical conductivity in these models. At very low frequencies, $\sigma(\omega)$ follows a simple Drude form. However,   for intermediate frequencies, $|\sigma(\omega)|$ has a power law fall off (with constant offset) and its phase is approximately constant. Remarkably, both the exponent of the power law and the phase are consistent with data taken on some cuprates and are robust against changing all parameters of our model.  
We do not have a deep understanding of why this is happening and it would clearly be of interest to find an analytic derivation of this result.

To get more insight into this result, there are a few generalizations that should be investigated. First, we have added the lattice in only one direction. It is natural to ask if the power or phase change if we do a more realistic calculation with a lattice in both spatial directions. We expect the answer will be no, but the calculation is considerably more difficult as it now requires solving nonlinear PDE's in three dimensions. Second, one can ask if the power and phase change if we consider a five dimensional bulk dual to a $3+1$ dimensional system. This is currently under investigation.

We have chosen to introduce a lattice by adding a periodic source for a neutral scalar operator. This induces a periodicity in the charge density, but the chemical potential $\mu$ is kept constant. One can alternatively model an ``ionic" lattice" by making $\mu$ itself a periodic function of $x$. (This is the approach taken in \cite{Hartnoll:2012rj,Flauger:2010tv,Maeda:2011pk}.) In this case, the dual gravitational description contains only Einstein-Maxwell theory. Work on this alternative lattice is underway and results will be reported elsewhere. Preliminary calculations indicate that the $\omega^{-2/3}$ scaling is seen for this lattice also.

We have taken just the first steps toward introducing lattice effects into holographic discussions of condensed matter systems. There are many future directions. For example, it would be interesting to determine the phonon spectrum.  Even though the lattice is fixed in the UV, the bulk bulk field it generates is dynamical and fluctuates. These fluctuations are governed by quasinormal modes.  

Another open problem is to add a charged scalar field to our gravitational model. Without the lattice, it is known that this theory has a second order phase transition at low temperature to a superconducting phase. It would be interesting to study the effect of the lattice on the transport properties of the superconductor.

Finally, one can add a probe fermion to the bulk. In the translationally invariant case, this was used to find Fermi surfaces. One can now study more realistic Fermi surfaces in the presence of a lattice. 

\appendix

\section{Appendix: Boundary Conditions}

In this Appendix we present the conditions on the perturbation at the horizon which implement ingoing wave boundary conditions. This can be easily achieved if one writes the background metric in terms of ingoing Eddington-Finkelstein coordinates, and require $h_{ab}$ to be a smooth tensor, at the horizon, in these coordinates. This induces the following near horizon behavior
\begin{align}
&\tilde{h}_{tt}(x,z)=(1-z)^{-\frac{i\,\omega}{P(1)}}[\tilde{h}_{tt}^{(0)}(x)+(1-z)\tilde{h}_{tt}^{(1)}(x)+\mathcal{O}((1-z))^2]\nonumber
\\
&\tilde{h}_{tx}(x,z)=(1-z)^{-\frac{i\,\omega}{P(1)}}[\tilde{h}_{tx}^{(0)}(x)+(1-z)\tilde{h}_{tx}^{(1)}(x)+\mathcal{O}((1-z))^2]\nonumber
\\
&\tilde{h}_{tz}(x,z)=(1-z)^{-1-\frac{i\,\omega}{P(1)}}[\tilde{h}_{tz}^{(0)}(x)+(1-z)\tilde{h}_{tz}^{(1)}(x)+\mathcal{O}((1-z))^2]\nonumber
\\
&\tilde{h}_{xx}(x,z)=(1-z)^{-\frac{i\,\omega}{P(1)}}[\tilde{h}_{xx}^{(0)}(x)+(1-z)\tilde{h}_{xx}^{(1)}(x)+\mathcal{O}((1-z))^2]\nonumber
\\
&\tilde{h}_{xz}(x,z)=(1-z)^{-1-\frac{i\,\omega}{P(1)}}[\tilde{h}_{xz}^{(0)}(x)+(1-z)\tilde{h}_{xz}^{(1)}(x)+\mathcal{O}((1-z))^2]\nonumber
\\
&\tilde{h}_{zz}(x,z)=(1-z)^{-2-\frac{i\,\omega}{P(1)}}[\tilde{h}_{zz}^{(0)}(x)+(1-z)\tilde{h}_{zz}^{(1)}(x)+\mathcal{O}((1-z))^2]\label{eqs:nearhexpansionper}\, .
\\
&\tilde{h}_{yy}(x,z)=(1-z)^{-\frac{i\,\omega}{P(1)}}[\tilde{h}_{yy}^{(0)}(x)+(1-z)\tilde{h}_{yy}^{(1)}(x)+\mathcal{O}((1-z))^2]\nonumber
\\
&\tilde{b}_{t}(x,z)=(1-z)^{-\frac{i\,\omega}{P(1)}}[\tilde{b}_{t}^{(0)}(x)+(1-z)\tilde{b}_{t}^{(1)}(x)+\mathcal{O}((1-z))^2]\nonumber
\\
&\tilde{b}_{x}(x,z)=(1-z)^{-\frac{i\,\omega}{P(1)}}[\tilde{b}_{x}^{(0)}(x)+(1-z)\tilde{b}_{x}^{(1)}(x)+\mathcal{O}((1-z))^2]\nonumber
\\
&\tilde{b}_{z}(x,z)=(1-z)^{-1-\frac{i\,\omega}{P(1)}}[\tilde{b}_{z}^{(0)}(x)+(1-z)\tilde{b}_{z}^{(1)}(x)+\mathcal{O}((1-z))^2]\nonumber
\\
&\tilde{\eta}(x,z)=(1-z)^{-\frac{i\,\omega}{P(1)}}[\tilde{\eta}^{(0)}(x)+(1-z)\tilde{\eta}^{(1)}(x)+\mathcal{O}((1-z))^2]\nonumber
\end{align}
where $P(1)$ is given in \eqn{p1}.
The boundary conditions are then fixed by inputing the above expansion into the equations of motion, and solving them in a $(1-z)$ expansion off the horizon. This expansion will fix some relations amongst the $\,^{(0)}$ coefficients. For instance, at the horizon regularity demands:
\begin{equation}
\tilde{b}_{z}^{(0)}(x)=\frac{L\, \tilde{b}_{t}^{(0)}(x)}{P(1)}.
\end{equation}
The remaining regularity conditions express how the $\,^{(1)}$ coefficients relate with some of the $\,^{(0)}$ coefficients and their tangential derivatives. These are far too cumbersome to be presented here.

\vskip 1cm
\centerline{\bf Acknowledgements}
\vskip 1cm

It is a pleasure to thank M. Fischer, S. Hartnoll, M. Metlitski, and D. Scalapino for discussions. This work began during the KITP program on Holographic Duality and Condensed Matter Physics. GH and DT are grateful to the KITP for hospitality during this time. This work was supported in part by the National Science Foundation under Grants No. PHY08-55415 and PHY11-25915 and ERC STG grant 279943, ``Strongly Coupled Systems".


\end{document}